\documentclass[twocolumn,showpacs,preprintnumbers,amsmath,amssymb,supers criptaddress]{revtex4}
\usepackage{graphicx}% Include figure files
\usepackage{dcolumn}% Align table columns on decimal point
\usepackage{bm}% bold math

\begin{document}

\draft
\title{Pulsed-gate measurements of the singlet-triplet relaxation time  in a two-electron double quantum dot}
\author{J.~R.~Petta}
\affiliation{Department of Physics, Harvard University, Cambridge,
MA  02138}
\author{A.~C.~Johnson}
\affiliation{Department of Physics, Harvard University, Cambridge,
MA  02138}
\author{A.~Yacoby}
\affiliation{Department of Physics, Harvard University, Cambridge,
MA  02138} \affiliation{Department of Condensed Matter Physics,
Weizmann Institute  of Science, Rehovot 76100 Israel}
\author{C.~M.~Marcus}
\affiliation{Department of Physics, Harvard University, Cambridge,
MA  02138}
\author{M.~P.~Hanson}
\affiliation{Materials Department, University of California, Santa
Barbara, CA 93106}
\author{A.~C.~Gossard}
\affiliation{Materials Department, University of California, Santa
Barbara, CA 93106}

\date{\today}

\begin{abstract} A pulsed-gate technique with charge sensing is used to  measure the singlet-triplet relaxation time for
nearly-degenerate spin states in a two-electron double quantum
dot. Transitions from the (1,1) charge occupancy state to the
(0,2) state,
 measured as a function of pulse cycle duration and magnetic field,  allow the (1,1) singlet-triplet relaxation time
($\geq $70 $\mu$s) and the (0,2) singlet-triplet splitting to be
measured. The use of charge sensing rather than current
measurement allows long relaxation times to be  readily probed.
\end{abstract}


\maketitle Semiconductor quantum dots are promising systems for
the  manipulation of electron spin because of the relative ease of
confining and measuring single electrons  \cite{Ciorga_PRB_2000}.
In order to make use of the
 spin degree of freedom as a holder of either classical or quantum  information, it is first necessary to understand and
characterize the mechanisms that lead to spin relaxation and
decoherence.

Previous studies of spin relaxation in quantum dots have focused
on  systems with large energy splittings of the relevant spin
 states, either singlet and triplet two-electron states, split by  $\sim$ 600 $\mu$eV
\cite{Fujisawa_Nature_2002}, or Zeeman states, split by $\geq$200
$\mu$eV
\cite{Hanson_PRL_2003,Elzerman_Nature_2004,Kroutvar_Nature_2004}.
In these cases, spin-orbit mediated processes are expected to
dominate spin relaxation
\cite{Khaetskii_PRB_2000,Golovach_PRL_2004}, while processes
involving hyperfine coupling of electron spin to nuclei are
relatively ineffective for splittings greater than the hyperfine
coupling scale $\sim$0.1 meV $/\sqrt{N_{E}}$$ \sim$1 $\mu$eV,
where $N_{E}$ is the effective number of nuclei interacting with
the electron spin \cite{Dobers_PRL_88,
Erlingsson_PRB_2001,Khaetskii_PRL_2002}. Spin relaxation between
nearly degenerate electronic states is particularly relevant to
the problem of controlled entanglement, as singlet-triplet
splitting goes to zero as the entangled spins become spatially
separated.

\begin{figure}[b]
\begin{center}
\includegraphics[width=7.5cm]{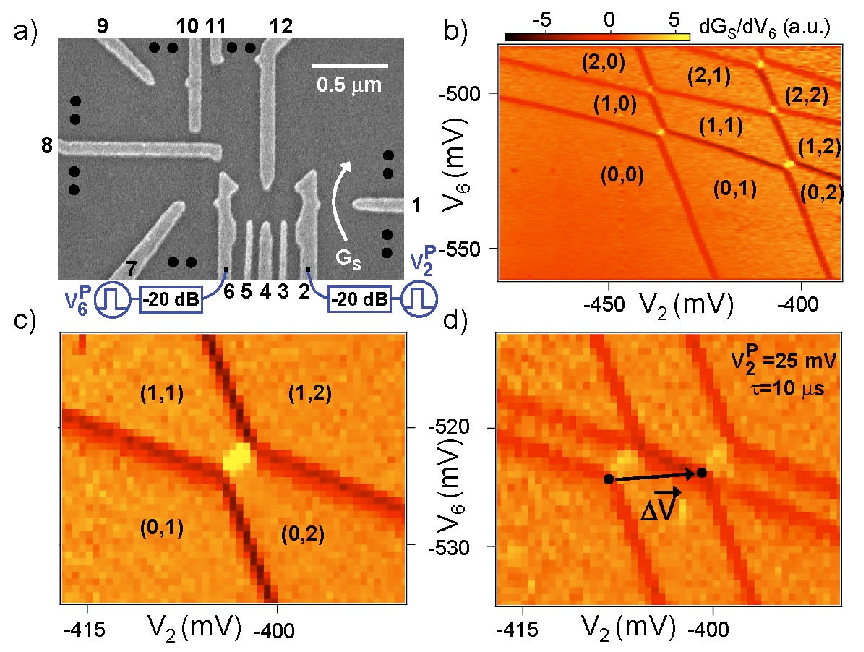}
\end{center}\vspace{-.1cm}
 \caption{(a) SEM image of a device identical in
design  to the one used in this experiment. Gates 2--6 and 12
define the double dot. $\bullet$ denotes an ohmic contact. (b)
Large-scale plot of $dG_{S}$/$dV_{6}$ as a function of $V_{2}$ and
$V_{6}$. Charge states are labelled $($$ M$,$N$$ )$, where $M$$
($$ N$$ )$ is the time averaged number of electrons on the left
(right) dot.  (c) Zoom in near the (1,1) to (0,2) charge
transition. (d) 25 mV pulses with a 50\% duty cycle and 10 $\mu$s
period are applied to the coax connected to gate 2. This results
in two copies of the charge stability diagram shifted  relative to
one another by $\Delta$$ \vec{V}$.} \vspace{-.1cm}
\end{figure}

In this Letter, we present pulsed-gate measurements of the
singlet-triplet relaxation times and splittings in a two-electron
double quantum dot. Spin states of the weakly-coupled  (1,1)
charge configuration (integer pairs specify the equilibrium charge
occupancy on the left and right dot) are nearly degenerate. A spin
selection rule impedes the transition from (1,1) spin triplets to
the (0,2) singlet. To probe spin  relaxation, the double dot is
initialized, then loaded into the (1,1) state, then placed in a
configuration where (0,2) is the preferred charge  distribution in
equilibrium. This transition probability is measured from a
calibrated charge sensing signal as a function of the period of
this three-step cycle. The resulting characteristic time of
$\sim$70 $\mu$s sets the time scale for of the  spin-blockaded
relaxation channels from (1,1)  to the singlet state of (0,2).
Measurements of the (1,1) to (0,2)  transition probability as a
function of detuning, $\epsilon$, show a strong dependence on
perpendicular magnetic field, $B_{\perp}$, reflecting the
singlet-triplet splitting of the (0,2) charge state, $J$ as
discussed  below.

Samples are fabricated from a GaAs/Al$_{0.3}$Ga$_{0.7}$As
heterostructure grown by molecular beam epitaxy. Electron beam
lithography and liftoff are used to create Ti/Au gates that
deplete a  100 nm deep two-dimensional electron gas with electron
density 2$\times$10$^{11}$cm$^{-2}$ and mobility
2$\times$10$^5$cm$^2$/V$\cdot$s. Gates 2--6 and 12 form the double
quantum dot [see Fig.\ 1(a)]. Gates 2 and 6 are connected via
bias tees to dc voltage sources and to pulse generators through
coax cables with $\sim$20 dB of inline attenuation
\cite{Tektronix}. A QPC charge detector is created by depleting
gate 1. Gates 7--11 are unused. The QPC conductance,  $G_S$, is
measured using standard lock-in amplifier techniques with a 1 nA
current bias at 93 Hz. The electron temperature,  T$_e$$ \sim$135
mK, was determined from Coulomb blockade peak widths. We  present
results from a single sample; a second sample with slightly larger
lithographic dimensions gave qualitatively similar  results
\cite{Petta_PRL_2004}.

QPC charge sensing is used to determine the absolute number of
electrons in the double dot. Figure 1(b) shows a large-scale
charge stability diagram for the double dot. As electrons  enter
or leave the double dot, or transfer from one dot to the other,
$G_S$ changes, resulting in sharp features in  $dG_S$/$dV_6$
(numerically differentiated)
\cite{Elzerman_PRB_2003,Petta_PRL_2004}. In the lower left corner
of  Fig.\ 1(b), the double dot is completely empty. As the gate
voltages are made more positive, electrons are added to the
double dot. We will focus on the two-electron regime near the
(1,1) to (0,2) charge transition [Fig.\ 1(c)]. In this  location,
charge transport in a double dot shows a striking asymmetry in
bias voltage due to spin selection rules (Pauli  blocking)
\cite{Ono_Science_2002,Johnson_submitted}. At forward bias,
transitions  from the (0,2)$_\textrm{S}$ to the (1,1) singlet
state (1,1)$_\textrm{S}$ are allowed. However, for reverse bias,
the  (1,1) to (0,2) transition can be blocked if the (1,1) state
forms a (1,1)$_\textrm{T}$ because the (0,2)$_\textrm{T}$  state
resides outside the transport window due to the large
singlet-triplet splitting in (0,2). This asymmetry results in
current rectification, which is used in present pulsed gate
measurements.

The double dot is electrostatically driven by applying pulses to
gates  2 and 6. Figure 1(d) shows a charge stability diagram
acquired with square pulses applied to gate 2 ($V_2^P$=25 mV,
50$\%$ duty cycle, period $\tau$=10 $\mu$s). This results in two
copies of the charge stability diagram, the right-most
(left-most) charge stability diagram reflects the ground state
charge configuration during the low (high) stage of the  pulse
sequence. The gate-voltage offset between the charge stability
diagrams, $\Delta \vec{V}$, is used to calibrate pulse
amplitudes. Additional calibrations are performed for gate 6,
which primarily shifts the honeycomb in the  vertical direction
(not shown). A linear combination of pulses on gates 2 and 6 can
be used to shift the stability diagram  in any direction in gate
space.

As a control, we compare a forward pulse sequence, where spin
selection  rules are expected to be important, with a reversed
pulse sequence that does not involve spin selective transitions.
The  forward pulse sequence (Fig.\ 2(a)) begins with the gates at
point E for 10\% of the period, emptying the second  electron from
the double dot, leaving the (0,1) charge state. The gates then
shift to the reset point R for the next  10\% of the period, which
initializes the system into the (1,1) configuration. The interdot
tunnel coupling, $t$, is  tuned with $t$$ \leq$$ k_BT$ so that the
(1,1) singlet-triplet splitting $j$$ \sim$$ 4t2$/$U$$ \ll$$ k_BT$,
where $U$ is the single dot charging energy. Due to this
degeneracy, and at low fields such that $|g|$$ \mu_B$$ B$$ <$$
k_BT$ ($|g|$$ \sim$0.44 for GaAs), we expect to load into
(1,1)$_\textrm{S}$ or any  of the three (1,1)$_\textrm{T}$ states
with equal probability. For the final 80\% of the period, the
gates are at the  measurement point M where (0,2)$_\textrm{S}$ is
the ground state. Figure 2(b) illustrates the possible (1,1) to
(0,2)  transitions. If the R step loads the (1,1)$_\textrm{S}$
state, tunneling to (0,2)$_\textrm{S}$ occurs on a  timescale
given by the interdot tunneling rate, $\Gamma(\epsilon)$ (We
estimate the slowest $\Gamma(\epsilon)$$
$(1 $\mu$s)$^{-1}$ from finite bias data
\cite{Johnson_submitted}). If the m$_\textrm{s}$=0 (1,1) triplet
state  (1,1)$_\textrm{T0}$ is loaded, it dephases into
(1,1)$_\textrm{S}$ on a timescale of $T_2$ (expected to be
$\leq$100 ns
\cite{Kikkawa_PRL_98,Khaetskii_PRL_2002,Merkulov_PRB_2002})
followed by a direct transition to (0,2)$_\textrm{S}$. About half
the  time the R step will load the m$_\textrm{s}$=1 (1,1) triplet
state (1,1)$_\textrm{T+}$ or the m$_\textrm{s}$=-1 (1,1)  triplet
state (1,1)$_\textrm{T-}$. At low $B_{\perp}$, (0,2)$_\textrm{T}$
is inaccessible, and a transition from  (1,1)$_\textrm{T+}$ or
(1,1)$_\textrm{T-}$ to (0,2) requires a spin flip and will be
blocked for times shorter than  the singlet-triplet relaxation
time $\tau_{ST}$.
\begin{figure}[t]
\begin{center}
\includegraphics[width=7cm]{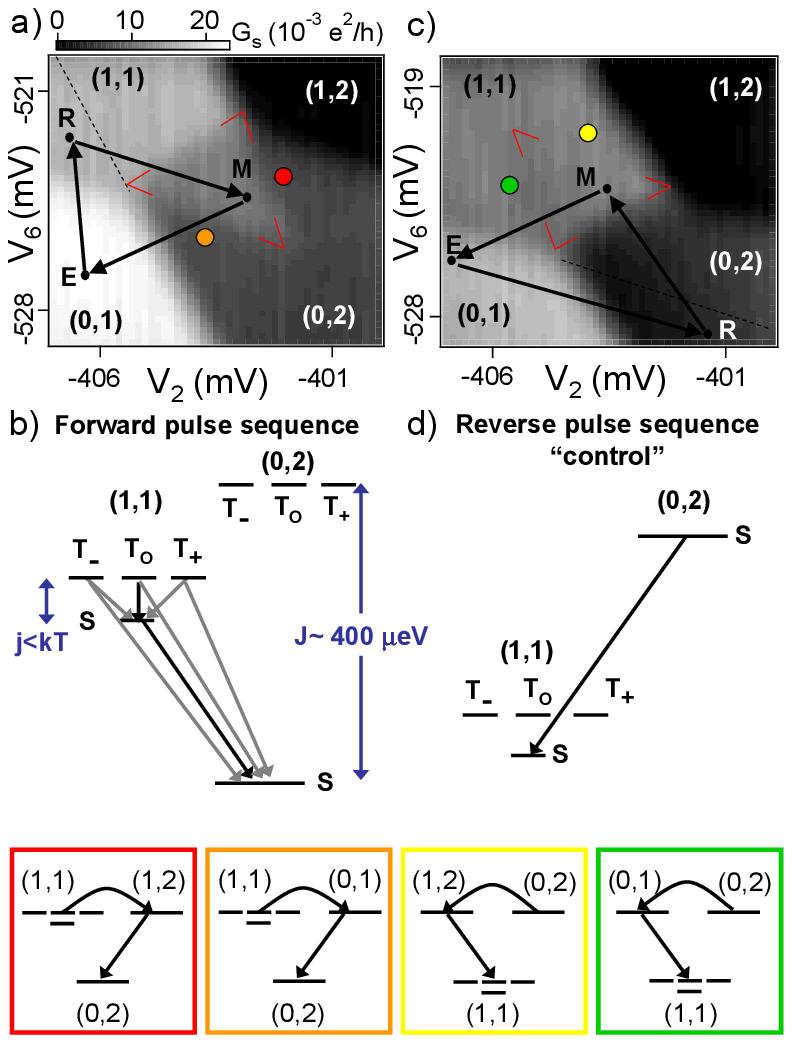}
\end{center}\vspace{-.2cm}
 \caption{(a) Sensor conductance, $G_S$, as a
function  of $V_2$ and $V_6$ while applying the forward pulse
sequence (see text) with $\tau$=10 $\mu$s and $B_{\perp}$=100 mT.
Spin-blocked transitions result in some (1,1) charge signal in the
(0,2) pulse triangle (bounded by the red lines).  Outside of the
pulse triangle it is possible to access other charge states. This
relaxes the spin blockade (see  color-coded level diagrams). (b)
Level diagram illustrating the possible transitions from (1,1) to
(0,2). Black arrows  indicate fast transitions, grey arrows
indicate spin blocked transitions. (c) $G_S$ as a function of
$V_2$ and $V_6$  while applying the reverse ``control" pulse
sequence with $\tau$=10 $\mu$s and $B_{\perp}$ =100 mT. The
(0,2)$_\textrm{S}$ to (1,1)$_\textrm{S}$ transition is not spin
blocked. As a result, there is no detectable pulse signal in the
pulse triangle (bounded by the red lines). (d) Level diagram
illustrating the (0,2) to (1,1) transition. A best-fit  plane has
been subtracted from the data in (a),(c) to remove signal from
direct gate to QPC coupling.}\vspace{-.2cm}
\end{figure}

In Fig.\ 2(a) the average charge sensor signal, $G_S$, is measured
as a  function of the dc gate voltages $V_2$ and $V_6$, while a
pulse sequence is repeated. This has the effect of translating
the points E, R, and M throughout the charge stability diagram,
keeping their relative positions constant. Because  most of the
time is spent at point M, the grayscale data primarily map out the
ground state population for this point, with  plateaus at $G_S$$
\sim$0.0, 6.0, 16, and 23 $\times$10$^{-3}$e$^2$/h indicating full
population of the (1,2), (0,2), (1,1), and (0,1) charge states
respectively. The pulse data differs from ground state  data only
when point M resides in the triangle defined by the (1,1) to (0,2)
ground state transition and the  extensions of the (1,1) to (0,1)
and (1,1) to (1,2) ground state transitions (bounded by the red
marks in Fig.\ 2(a)). Within this  ``pulse triangle" transitions
from (1,1) to (0,2) may be blocked as described above, and the
charge sensor  registers a conductance intermediate between the
(1,1) and (0,2) plateaus. If M moves above the pulse triangle (red
dot in  Fig.\ 2(a)), the (1,1) to (0,2) transition can occur
sequentially via (1,2) with no interdot tunneling: a new electron
enters the right dot, then the electron in the left dot leaves.
Likewise, if M moves below the pulse triangle (orange dot  in
Fig.\ 2(a)) the transition can occur via (0,1): the left-dot
electron leaves,  then a new electron enters the right dot. By
similar logic, point R must be to the left of the (0,1) to (0,2)
transition extension (dotted line in Fig.\ 2(a)) to avoid
resetting through (0,2) and preferentially loading
(1,1)$_\textrm{S}$. Figure 2(a) shows a signal of
~11$\times$10$^{-3}$e$^2$/h in the pulse triangle for $\tau$=10
$\mu$s,  which indicates that approximately 50\% of the time the
dots remain in (1,1) even though (0,2) is the ground state.  This
is direct evidence of spin-blocked (1,1) to (0,2) transitions.

If the the pulse sequence is reversed, so that the reset position
R occurs in (0,2), where only the singlet state is accessible, and
M occurs in (1,1), tunneling from R to  M should always proceed on
a time scale set by the interdot tunnel  coupling, since the
(0,2)$_\textrm{S}$ to (1,1)$_\textrm{S}$ transition is not spin
blocked. As anticipated, no  signal is seen in the pulse triangle
for this reversed ``control" sequence (Fig.\  2(c)).

Spin selectivity of the forward pulse sequence in Fig.\ 3 can be
used to measure $J$ as a function of $B_\perp$
\cite{Ashoori_PRL_1993}. Figure 3(a) shows $G_S$  as a function of
$V_2$ and $V_6$ while applying the forward pulse sequence with
$B_\perp$$ =$1.2 T and $\tau$$ =$10 $\mu$s. For these data,
(0,2)$_\textrm{T}$ resides outside of the  pulse triangle ($J$$
$$
E_M$, the mutual charging energy) and the (1,1)$_\textrm{T}$ to
(0,2)  transitions are spin blocked. For $B_\perp$$ =$1.4 T [Fig.\
3(b)] the (0,2)$_\textrm{T}$ state is low enough in  energy that
the (1,1)$_\textrm{T}$ states can directly tunnel to the
(0,2)$_\textrm{T}$ manifold at high detunings.  Now (1,1) to (0,2)
tunneling can proceed, and there is no longer a (1,1) charge
signal in the (0,2) region of the  pulse triangle at high
detuning. This cuts off the tip of the pulse triangle. The
spin-blocked region continues to shrink  as $B_\perp$ is
increased. From these data, we find $J$$ \sim$340, 280, and 180
$\mu$eV for $B_\perp$=1.4, 1.6, and 1.8 T,  respectively
\cite{lever_arm}.

\begin{figure}[t]
\begin{center}
\includegraphics[width=7.2cm]{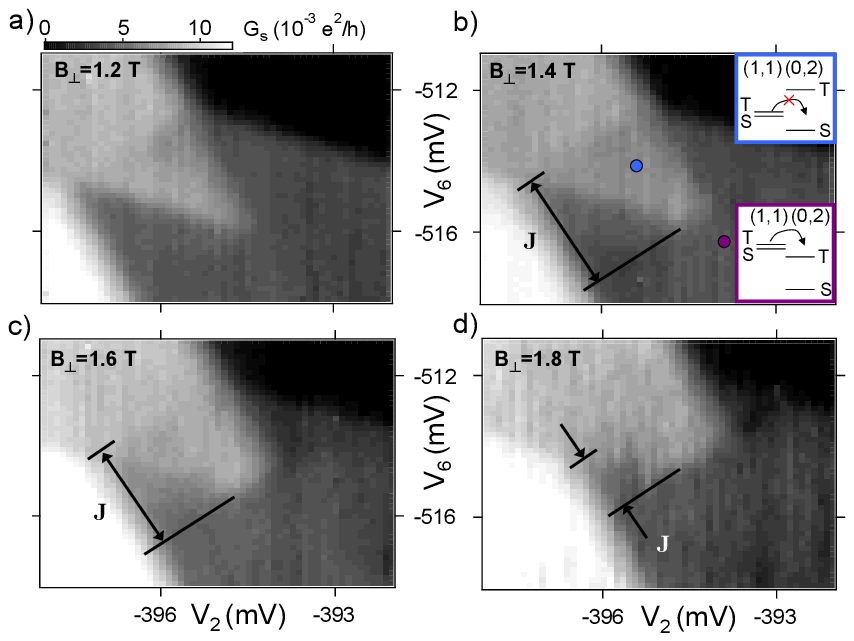}
\end{center}\vspace{-.2cm}
 \caption{Forward pulse sequence results with
$\tau$=10  $\mu$s. (a--d) $G_{s}$ as a function of $V_2$ and $V_6$
for increasing $B_{\perp}$. $B_{\perp}$ reduces $J$.  Eventually
(0,2)$_\textrm{T}$ is lowered into the ``M" pulse window [inset of
(b)]. At this point, (1,1)$_\textrm{T}$ to  (0,2)$_\textrm{T}$
transitions are energetically possible and the transition from the
(1,1) to (0,2) charge state is no  longer ``spin blocked". This
cuts off the tip of the pulse triangle in (0,2), see (b). A
best-fit plane has been  subtracted from the data in
(a)--(d).}\vspace{-.2cm}
\end{figure}

The time dependence of the charge sensing signal can be
investigated by  varying $\tau$, the overall period of the cycle.
Figure 4(a) shows $G_S$ as a function of $V_2$ and $V_6$ acquired
using the forward pulse sequence  with $\tau$=8 $\mu$s at
$B_{\perp}$=100 mT. A clear pulse signal is observed in the pulse
triangle. As $\tau$ is  increased, the pulse signal decreases as
shown in (b--c). $G_S$ is measured inside the pulse triangle
($V_2$,$V_6$ held  fixed at -403,-523.8 mV, respectively) and is
plotted as a function of $\tau$ in Fig.\ 4(d). In (1,1), $G_S$$
\sim $20$\times$10$^{-3}$e$^2$/h, whereas outside the pulse
triangle in  (0,2), $G_S$$ \sim $10$\times$10$^{-3}$e$^2$/h. For
small $\tau$, $G_S$$ \sim $15$\times$10$^{-3}$e$^2$/h in the pulse
triangle. At long $\tau$,  $G_S$ approaches
10$\times$10$^{-3}$e$^2$/h in the pulse triangle, which indicates
complete transfer from the (1,1) to  (0,2) charge state.

These data are consistent with spin-blocked transitions preventing
the  (1,1) to (0,2) charge transition. Approximately 50$\%$ of the
time, the (1,1) R pulse loads into either  (1,1)$_\textrm{T+}$ or
(1,1)$_\textrm{T-}$. These states may relax into the
(1,1)$_\textrm{S}$ state and then tunnel to  (0,2)$_\textrm{S}$ on
a timescale set by $\tau_{ST}$. For $\tau$$ \ll$$ \tau_{ST}$, we
expect that 50$\%$ of the (1,1) to (0,2) transitions  will be
spin-blocked, resulting in $\sim$50$\%$ (1,1) pulse signal in the
(0,2) pulse triangle. For $\tau$$
$$
\tau_{ST}$, the (1,1)$_\textrm{T+}$ and (1,1)$_\textrm{T-}$ states
have  ample time to relax to (1,1)$_\textrm{S}$, after which a
(1,1)$_\textrm{S}$ to (0,2)$_\textrm{S}$ transition can take
place. Thus, for $\tau$ long compared to $\tau$$ _{ST}$, the pulse
signal to approaches the (0,2) level, indicating full  transfer
from (1,1) to (0,2). In the intermediate $\tau$ regime, the sensor
signal due to spin-blocked  transitions decays as a function of
time on a timescale that is characteristic of $\tau$$ _{ST}$.

The experimental data in Fig.\ 4(d) are fit assuming exponential
singlet-triplet relaxation. Due to the slow measurement rate of
the charge sensor ($\sim$100 ms), $G_S$ is proportional to the
time-averaged occupation of the left dot.  Modeling an exponential
decay of the  sensing signal weighted over the 80\% of the cycle
corresponding to the (0,2) measurement gives  $G_S(\tau)$$ =$$ A$$
+$$ B$$ (\tau_{ST}/\tau)$$ (1-e^{-0.8\tau/\tau_{ST}})$, where A is
the conductance asymptote at  long times (full occupation of the
(0,2) state) and B is the additional conductance in a short pulse
due to the blocked  states (approximately 50\% of the (0,2) to
(1,1) step height).  The best fit to the data in Fig.\ 4(d) gives
A=0.009  e$^2$/h and B=0.007 e$^2$/h, consistent with these
expectations. In the center of the pulse triangle the best-fit
$\tau_{ST}$ reaches a maximum of $\tau_{ST}$$ =$70$\pm$10 $\mu$s.
Near the (1,1) to (0,2) transition, $\tau_{ST}$  decreases, to
20$\pm$5 $\mu$s at $V_2$=-403.8 mV and $V_6$=-523.0 mV. Closer to
the tip of the pulse triangle,  $\tau_{ST}$ decreases due to
thermally activated exchange with the leads (See Fig.\ 2, red and
orange diagrams), thus the 70  $\mu$s relaxation time represents a
lower bound on the spin relaxation time within the (1,1) manifold
\cite{interdot}.

\begin{figure}[t]
\begin{center}
\includegraphics[width=7.2cm]{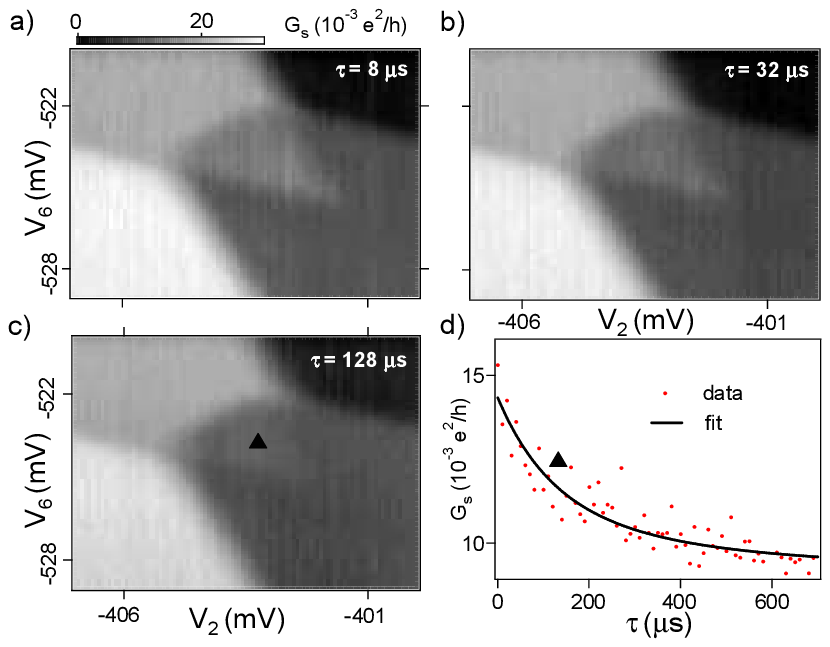}
\end{center}\vspace{-.2cm}
 \caption{Forward pulse sequence results with
$B_{\perp}$=100 mT. (a)--(c) $G_{S}$ as a function of $V_2$ and
$V_6$ for increasing pulse periods, $\tau$. For longer  periods,
singlet-triplet relaxation occurs and the (1,1) to (0,2)
transition proceeds, reducing the signal in the (0,2)  pulse
triangle. (d) $G_{s}$ with $V_2$=-403.0 mV, $V_6$=-523.8 mV (gate
voltage position indicated by the black triangle  in (c)) measured
as a function of $\tau$. $G_{s}$=10$\times$10$^{-3}$e$^2$/h in the
(0,2) charge state, while  $G_{s}$=20$\times$10$^{-3}$e$^2$/h in
the (1,1) charge state. For short $\tau$, the (1,1) to (0,2)
transition is  blocked approximately half of the time, resulting
in a pulse signal of 15$\times$10$^{-3}$e$^2$/h in the (0,2) pulse
triangle.  The (1,1) to (0,2) transition probability increases
(sensor signal decreases) with $\tau$ due to spin-relaxation  with
a characteristic time scale of 70$\pm$10 $\mu$s. A best-fit plane
has been subtracted from the data in  (a)--(c).}
\end{figure}

In GaAs quantum dots, spin-orbit and hyperfine interactions are
expected to be the dominant mechanisms for spin relaxation. The
spin-orbit contribution to T$_1$ depends strongly on  $B$ and
theory predicts $T_1$$ =$10 ms for $B$$ =$1 T
\cite{Golovach_PRL_2004}. The hyperfine spin relaxation rate is
$\Gamma_{HF}$$ =$$ \Gamma_{ph}$$ \left(E_n \over j\right)2
\frac{\gamma_{n} G_{C}}{(2I)^2N_{E}}$ where $\Gamma_{ph}$$
\approx$4$\times$10$^7$s$^{-1}$ reflects the electron-phonon
coupling,  $E_n$=0.135 meV is the nuclear energy for a fully
polarized system, $G_{C}$=1.25 is derived from the correlation of
the  nuclear spins, $\gamma_{n}$=0.045 depends on wavefunction
overlap, the nuclear spin I=3/2, and $N_{E}$$ \sim$10$^4$
\cite{Erlingsson_PRB_2001}. Our sample is tuned with $j$$ \sim$$
4t2$/$U$$ \ll$$ k_BT$. In this case, the Zeeman splitting between
the $m_s$=0 state and  $m_s$=$\pm$1 states sets the relevant
energy scale. Using $j$$ \sim$2.5$\mu$eV gives $1/\Gamma_{HF}$=15
$\mu$s, roughly consistent  with the experimental results. The
relaxation times found here can be compared  with those obtained
by Fujisawa \textit{et al.}\, who measured $\tau_{ST}$=200 $\mu$s
in a vertical dot \cite{Fujisawa_Nature_2002}. In those
measurements, the sizable singlet-triplet splitting implies that
spin-orbit processes rather than nuclear processes dominant, in
contrast to the present experiments.

\begin{acknowledgments} We acknowledge useful discussions with Jacob  Taylor, Hans-Andreas Engel, and Mikhail Lukin. This
work was supported by the ARO under DAAD55-98-1-0270 and
DAAD19-02-1-0070, DARPA under the QuIST program, the NSF under
DMR-0072777 and the Harvard NSEC.
\end{acknowledgments}


\begin{thebibliography}{99}
\bibitem{Tektronix} We use Anritsu K251 bias tees. Tektronix AWG520  pulse generators are used for high speed
manipulation of the gate voltages.
\bibitem{lever_arm} The conversion from gate voltage to energy is  determined by measuring finite bias triangles.
\bibitem{interdot} $\Gamma({\epsilon})$ is set to be much faster than  the characteristic relaxation times that we
measure in this experiment. $\Gamma({\epsilon})$ varies strongly
with  $\epsilon$ (see ref.\ 13). The fact that $\tau_{ST}$ only
varies by roughly a factor of three in the pulse  triangle also
indicates that the (1,1)-(0,2) transitions are spin-blocked and
not limited by a slow  $\Gamma({\epsilon})$.
\end{thebibliography}
\end{document}